\documentclass[10pt]{article}
\usepackage{amssymb,amsmath,amsfonts,mathrsfs,enumerate,wrapfig}
\usepackage{graphicx,rotate,multicol}
\usepackage{bm}
\usepackage{cite}
\usepackage[colorlinks=true,
            linkcolor=red,
            urlcolor=blue,
            citecolor=blue]{hyperref}

\usepackage{color}
\usepackage{tikz}

\def\tb{\tan\beta}

\def \order(#1){{\cal O} \left(#1 \right)}
\long\def\rpl#1!!#2!!{\textcolor{red}{#1} \textcolor{blue}{#2}}

\evensidemargin=\oddsidemargin
\def\Eqn#1{Eq.\ (\ref{#1})}

\allowdisplaybreaks

\def\bar {\overline}

\def\be {\begin{equation}}
\def\ee {\end{equation}}
\def\beq {\begin{equation}}
\def\eeq {\end{equation}}
\def\bea {\begin{eqnarray}}
\def\eea {\end{eqnarray}}

\def\beq{\begin{equation}}
\def\eeq{\end{equation}}
\def\barr{\begin{array}}
\def\earr{\end{array}}

\def\opcit(#1){ {\em op. cit.}, #1}

\def\issue(#1,#2,#3){#1, #2 (#3)} 

\def\APP(#1,#2,#3){Acta Phys.\ Polon.\ \issue(#1,#2,#3)}
\def\ARNPS(#1,#2,#3){Ann.\ Rev.\ Nucl.\ Part.\ Sci.\ \issue(#1,#2,#3)}
\def\CPC(#1,#2,#3){Comp.\ Phys.\ Comm.\ \issue(#1,#2,#3)}
\def\CIP(#1,#2,#3){Comput.\ Phys.\ \issue(#1,#2,#3)}
\def\EPJC(#1,#2,#3){Eur.\ Phys.\ J.\ C\ \issue(#1,#2,#3)}
\def\EPJD(#1,#2,#3){Eur.\ Phys.\ J. Direct\ C\ \issue(#1,#2,#3)}
\def\IEEETNS(#1,#2,#3){IEEE Trans.\ Nucl.\ Sci.\ \issue(#1,#2,#3)}
\def\IJMP(#1,#2,#3){Int.\ J.\ Mod.\ Phys. \issue(#1,#2,#3)}
\def\JHEP(#1,#2,#3){J.\ High Energy Physics \issue(#1,#2,#3)}
\def\JPG(#1,#2,#3){J.\ Phys.\ G \issue(#1,#2,#3)}
\def\MPL(#1,#2,#3){Mod.\ Phys.\ Lett.\ \issue(#1,#2,#3)}
\def\NP(#1,#2,#3){Nucl.\ Phys.\ \issue(#1,#2,#3)}
\def\NIM(#1,#2,#3){Nucl.\ Instrum.\ Meth.\ \issue(#1,#2,#3)}
\def\PL(#1,#2,#3){Phys.\ Lett.\ \issue(#1,#2,#3)}
\def\PRD(#1,#2,#3){Phys.\ Rev.\ D \issue(#1,#2,#3)}
\def\PRL(#1,#2,#3){Phys.\ Rev.\ Lett.\ \issue(#1,#2,#3)}
\def\SJNP(#1,#2,#3){Sov.\ J. Nucl.\ Phys.\ \issue(#1,#2,#3)}
\def\ZPC(#1,#2,#3){Zeit.\ Phys.\ C \issue(#1,#2,#3)}

\usepackage[normalem]{ulem}

\definecolor{darkgreen}{cmyk}{1,0,1,0.4}
\definecolor{pink}{cmyk}{0.4,1,0.3,0}

\begin{document}

\renewcommand*{\thefootnote}{\fnsymbol{footnote}}

\begin{center}
 {\Large\bf{
Two more hidden scalars around 125 GeV and $\bm{h\to\mu\tau}$}}\\
\vspace*{0.5cm}
{  Dipankar Das$^{a,}$\footnote{d.das@saha.ac.in}, ~Anirban Kundu$^{b,}$\footnote{anirban.kundu.cu@gmail.com}} \\

	\vspace{10pt} {\small } {\em $^a$Theory Physics Division, Saha Institute of Nuclear
		Physics, \\ 1/AF Bidhan Nagar, Kolkata 700064, India \\ \vspace{3mm}
		$^b$Department of Physics, University of Calcutta, \\
		92 Acharya Prafulla Chandra Road, Kolkata 700009, India}

\end{center}
\begin{abstract}

We show that the $2.4\sigma$ signal of the leptonic flavor violating (LFV) Higgs 
boson decay $h\to\mu\tau$, as observed by the CMS collaboration recently, 
can be explained by a certain class of two-Higgs doublet models that allow 
controllable flavor-changing neutral current with minimal number of free 
parameters. We postulate that (i) the alignment limit is maintained, which means 
the lightest neutral scalar ($h$) has identical couplings to that of the Standard Model
Higgs boson and (ii) the signal comes from two other neutral scalars,
the CP-even $H$ and the CP-odd $A$, almost degenerate 
with $h$ at 125 GeV. We also show that (i) it is 
entirely possible that these scalars are hidden, apart from this LFV signal; 
(ii) the signal strengths of $b\bar{b}$, $\tau^+\tau^-$
and $\gamma\gamma$ around 125 GeV put severe constraints on the parameter space of such models; 
(iii) the constraint is further enhanced by the non-observation of processes like
$\mu\to e\gamma$, and we predict that the branching ratio of $\mu\to e\gamma$ cannot be 
even an order below the present experimental limit, highlighting the role it plays in 
forcing $H$ and $A$ to be near-degenerate; 
(iv) an enhancement in the $\tau^+\tau^-$ production cross-section at around 125  GeV is 
expected in the gluon fusion channel, 
and should be observed during the next run of the LHC; (v) the branching ratio in the 
$e\tau$ channel is enhanced and is expected to be at least about $2\%$. 
The constrained parameter space 
and minimum number of free parameters, along with such strong predictions, 
make this model easily testable and falsifiable.

\end{abstract}

\date{\today} 


\setcounter{footnote}{0}
\renewcommand*{\thefootnote}{\arabic{footnote}}

\section{Introduction}

Leptonic flavor violating (LFV) processes do not take place in the Standard Model 
(SM); even with massive neutrinos, they are expected to be unobservably small, 
because the amplitudes are controlled by the tiny neutrino masses. Thus, observation
of any LFV process is a smoking gun signal for physics beyond the SM \cite{Vicente:2014mya}. No such 
signal has been observed so far in processes like $\ell_1\to \ell_2\gamma$, 
$\ell_1\to \ell_2 \bar{\ell_3}\ell_4$, $\ell_1 \to \ell_2 M$, $M_1\to M_2 \ell_1
\bar{\ell_2}$, where $\ell$ and $M$ stand for a generic lepton and meson, and the 
indices are arranged in such a way that the processes are both LFV and kinematically 
allowed. 

Recently, the CMS collaboration\cite{Khachatryan:2015kon} found a $2.4\sigma$ signal in the Higgs 
boson decay channel $h\to\mu\tau$. This excess has been observed in both leptonic and 
hadronic final state channels of $\tau$; the weighted average is
\be
{\rm Br}(h\to\mu\tau) = 0.84^{+0.39}_{-0.37}\%\,,
\label{cms-data}
\ee
and the upper limit at 95\% confidence limit (CL) is ${\rm Br}(h\to\mu\tau) < 1.51\%$.

There have been several attempts in the literature to explain this signal by introducing 
LFV couplings of the Higgs boson. This can be achieved by an extension of the scalar sector
(and if necessary, the gauge and fermion sectors too), with some discrete (like $S_4$ \cite{Campos:2014zaa} 
or $A_4$ \cite{Heeck:2014qea}) or continuous (like gauged $L_\mu-L_\tau$ \cite{Crivellin:2015mga} 
or an $U(1)'$ with two scalar doublets \cite{Crivellin:2015lwa}) symmetries,
or supersymmetric Froggatt-Nielsen mechanism \cite{Dery:2014kxa}. All these extensions necessarily 
introduce a number of new arbitrary parameters in the model. It was found \cite{Dorsner:2015mja} 
that an extension of the scalar sector is imperative to explain the anomaly. As a concrete example, 
phenomenology of the type-III two-Higgs doublet model (2HDM) was considered in detail 
in \cite{Crivellin:2015mga,Crivellin:2015lwa,Sierra:2014nqa,Dorsner:2015mja,Omura:2015nja}, 
including some predictions for the model.
The general feature of all these models is to predict at least one new scalar with LFV couplings 
that mixes with the SM doublet $\Phi$ and the resultant mass eigenstate, which is dominantly
$\Phi$ with a small admixture of the LFV scalar, showing LFV signals while being in conformity 
with the SM for flavor-conserving decay channels. 

In this paper, we would like to explore the consequences of a particular class 
of 2HDM \cite{Branco:2011iw}, first proposed by Branco, Grimus, and Lavoura (BGL) \cite{Branco:1996bq}, 
which has tree-level flavor-changing neutral current (FCNC) interactions, appropriately suppressed 
by the entries of the quark or neutrino mixing matrix elements. The number of new parameters introduced
in this class of models is minimal. In a certain limit (called the alignment limit) 
motivated by the LHC data, a particular type 
of the BGL model can not only explain the $h\to\mu\tau$ signal, but also turns out to be 
extremely predictive. While a slight deviation from the alignment limit is still possible, 
resulting in the small-admixture explanation mentioned in the previous paragraph, there is 
another much more interesting possibility that we would like to explore.

We speculate that in the alignment limit, the second CP-even neutral scalar $H$ is 
almost degenerate with the Higgs boson $h$;  
so that the LFV signal comes from $H$ and not $h$,
which can decay in flavor-blind channels. But that is not the end of the story; $H$ and the 
CP-odd neutral scalar $A$ both contribute to $\mu\to e\gamma$ (and other LFV processes as well,
but this is most tightly constrained). If we want to keep Br($\mu\to e\gamma$) within the 
experimental limit, $H$ and $A$ must also be near-degenerate. Thus, there are three neutral scalars
sitting around 125 GeV, among which one is identical with the SM Higgs boson and the other two can have 
both flavor-conserving and flavor-violating couplings. We will discuss all the constraints 
on the parameter space, and show how one can successfully hide these two new scalars from 
the current LHC data. Concept of such degenerate Higgs bosons and their search strategies have 
been discussed in the literature \cite{Gunion:2012he,Ferreira:2012nv}. 
One may note that signatures of light scalars of this model in the LHC experiments may not yet be 
observable \cite{Bhattacharyya:2014nja}.

If this model is true, we have a few tangible predictions. On the theoretical side, 
the ratio of two vacuum expectation values (VEV), commonly parametrized by $\tan\beta$, must lie in a narrow range, 
something like $0.4 < \tan\beta < 2.8$. This range can further be narrowed down
with a more precise measurement of the neutrino mixing matrix. 
Another crucial input is the $h\to\tau\tau$ signal strength in the gluon fusion (ggF) production channel, and, as we will show, may 
point to an even narrower range of $\tan\beta$ centered around $\tan\beta=1$. 
On the experimental side, 
the next generation experiments 
looking for $\mu\to e\gamma$ should be able to see it, as the rate must be close to the upper bound 
unless there is a fine tuned cancellation. Apart from that, there will be a significant excess 
in $h\to\tau\tau$ (which, by default, also includes $H\to\tau\tau$ and $A\to\tau\tau$) in the 
gluon fusion channel, as the other scalars do not have any gauge couplings and hence cannot be produced in the 
vector boson fusion (VBF) channel. Precision study of the Higgs boson, either at the upgraded run 
of the LHC or at some future $e^+e^-$ collider, will certainly be able to test this model. For a current status 
of precision studies in the context of 2HDMs, we refer the reader to Ref.\ \cite{Cheung:2013rva}.

\section{Formalism}

The scalar potential part of the BGL model \cite{Branco:1996bq, Bhattacharyya:2013rya} is like 
the other canonical 2HDMs with $U(1)$ symmetry, for whose formalism we will not go into detail. 
A soft breaking term is introduced in the scalar potential to prevent the appearance of a 
massless pseudoscalar \cite{Bhattacharyya:2013rya}. 
For the Yukawa part, we will follow the notations of Ref.\ \cite{Branco:1996bq} as much as possible. 
The Yukawa part of the Lagrangian is given by
\be
 {\cal L}_Y = - \sum_{j=1}^2\sum_{\alpha,\beta=1}^3 \left(
 (\Gamma^q_j)_{\alpha\beta} \bar{Q'}_\alpha \Phi_j d'_\beta + 
 (\Delta^q_j)_{\alpha\beta} \bar{Q'}_\alpha (i\sigma_2\Phi_j^\ast) u'_\beta +
 (\Gamma^\ell_j)_{\alpha\beta} \bar{L'}_\alpha \Phi_j e'_\beta + 
 (\Delta^\ell_j)_{\alpha\beta} \bar{L'}_\alpha (i\sigma_2\Phi_j^\ast) \nu'_\beta 
 + {\rm h.c.} \right)\,,
\ee
where $\Phi_1$ and $\Phi_2$ are the two Higgs doublets, $\Gamma^q (\Gamma^\ell)$ and $\Delta^q (\Delta^\ell)$ 
are the Yukawa coupling matrices in the quark (lepton) sector, and $u'$, $d'$, $e'$, $\nu'$, $Q'$ and $L'$ 
stand for right-chiral up-type, down-type, charged lepton, neutrino,\footnote{We 
introduce right-handed neutrinos and assume, for simplicity, that the neutrinos are pure Dirac particles.}
and the left-chiral SU(2) doublet quark and lepton fields respectively. The primes indicate that these fields are in the gauge basis;
the mass basis fields will be denoted without primes. 
The generation indices are denoted by $\alpha$ and $\beta$.
Both $\Gamma^{q(\ell)}$ and $\Delta^{q(\ell)}$ are $3\times 3$ matrices. 

Now we impose the following BGL symmetry on the Lagrangian\cite{Branco:1996bq}:
\begin{eqnarray}
{\cal S}~: ~~~ Q'_k\to e^{i\theta} Q'_k,~ u'_k\to e^{2i\theta} u'_k,~ \Phi_2\to e^{i\theta} \Phi_2
~~~~ (k=~{\rm either}~1,~2~ {\rm or}~3) \,.
\label{BGL symmetry}
\end{eqnarray}
All the fields except those which appear above, remain unaffected under ${\cal S}$. Note that \Eqn{BGL symmetry}
violates lepton flavor universality by construction because we have singled out any one of the up-type quark 
fields and labeled it as $k$. Which up-type quark is labeled as $k$ will lead to different models within the BGL 
class. Such a symmetry leads to FCNC in the down-quark sector. One can put the nontrivial transformation to 
$d'_k$  instead of $u'_k$ in \Eqn{BGL symmetry}, which leads to FCNC in the up-type sector. Since the FCNC 
constraints are much tighter for down-type quarks, the former class of BGL models are more predictive 
and we will focus on them only.
We shall call our model u-, c- or t-type in accordance with $k=$1, 2 or 3 respectively in \Eqn{BGL symmetry}.

One can think of such an abelian symmetry for the lepton sector too. Note that $k$ for leptons need not be the 
same as $k$ for quarks. In fact, one can even have tree-level FCNC for up-type quarks and charged leptons. This is 
something that the reader should keep in mind when we discuss the possible constraints on these models.

The CP-even\footnote{We will 
assume all terms in the scalar potential to be real, so that the mass eigenstates 
are CP-eigenstates too.} neutral components of $\Phi_1$ and $\Phi_2$ mix with each 
other to give the mass eigenstates $h$ and $H$, and the mixing angle 
is usually denoted by $\alpha$.
An intermediate basis $\{H^0,R\}$ (not the mass basis in general) can be obtained from the  
$\{\Phi_1,\Phi_2\}$ basis by a rotation through $\beta \equiv \tan^{-1}(v_2/v_1)$ 
with the property that the state $H^0$ possesses exact
SM-like couplings with the fermions and the vector bosons.
The connection between these two bases is given by
\be
H^0 = \cos(\beta-\alpha)H + \sin(\beta-\alpha)h \,, ~~~~
R = -\sin(\beta-\alpha)H + \cos(\beta-\alpha)h \,.
\ee
Clearly, if we require that the lighter CP-even mass eigenstate, $h$, should posses SM-like couplings then
we are led to the alignment limit\footnote{If we require the heavier scalar, $H$, to be identical to $H^0$ then 
we are led to $\cos(\beta-\alpha)\approx 1$. Since we will be assuming $h$ and $H$ to be quasi-degenerate, 
this limit is also a possibility. Whatever we comment here in the context of the limit $\sin(\beta-\alpha) =1$ 
is equally valid also for the limit $\cos(\beta-\alpha)= 1$. More accurate measurements for the masses 
will be necessary to pinpoint the mass hierarchy and thereby to distinguish between these two limits.}
$\sin(\beta-\alpha) \approx 1$.
This is the limit favored by the current 2HDM fits 
\cite{Coleppa:2013dya,Chen:2013rba,Craig:2013hca, Eberhardt:2013uba,Dumont:2014wha,Dumont:2014kna,Bernon:2014vta,Chowdhury:2015yja}. From now on, we will talk only about the states $h$, $H$, and $A$.

The symmetry ${\cal S}$ of \Eqn{BGL symmetry} leads to a very specific texture of the 
Yukawa matrices \cite{Branco:1996bq, Botella:2014ska} which
leads to the following Yukawa Lagrangian in the alignment limit:
\begin{eqnarray}
{\cal L}_{Y} &=&
-\frac{1}{v}{h}\left[{\bar{d}}D_d{d}+{\bar{u}}D_u{u}\right]
- \frac{1}{v}{H}\left[{\bar{d}}(N_dP_R+N_d^\dagger
  P_L){d}+{\bar{u}}(N_uP_R+N_u^\dagger P_L){u}\right] \nonumber \\
&& +\frac{i}{v}{A}\left[{\bar{d}}(N_dP_R-N_d^\dagger P_L){d}-{\bar{u}}
(N_uP_R-N_u^\dagger P_L){u}\right]
+\left\{\frac{\sqrt{2}}{v}H^+ \bar{u}\left(VN_dP_R-N_u^\dagger 
VP_L\right){d} + {\rm h.c.}\right\} \,,
\label{full-yukawa}
\end{eqnarray}
where we have suppressed the generation indices, and the unprimed fermion fields are in the mass basis. 
$D_d$ and $D_u$ are diagonal matrices with the corresponding Yukawa couplings 
as the diagonal entries, so that the couplings of $H^0$ with the quarks are identical with the 
SM ones. 
The coupling matrices $N_u$ and $N_d$ 
are given by
\begin{eqnarray}
&& N_u^{\bf u} = {\rm diag}\{-m_u \cot\beta\,, m_c\tan\beta\,,
m_t\tan\beta\}\,, \ \ 
(N_d)_{ij}^{\bf u} = \tan\beta~m_i \delta_{ij}
-(\tan\beta+\cot\beta)V_{ui}^* V_{uj} m_j\, , \nonumber\\
&& N_u^{\bf c} = {\rm diag}\{m_u \tan\beta\,, -m_c\cot\beta\,,
m_t\tan\beta \}\,, \ \ 
(N_d)_{ij}^{\bf c} = \tan\beta~m_i \delta_{ij}
-(\tan\beta+\cot\beta)V_{ci}^* V_{cj} m_j\, , \nonumber\\
&&N_u^{\bf t} = {\rm diag}\{m_u \tan\beta\,, m_c\tan\beta\,,
-m_t\cot\beta \}\,, \ \ 
(N_d)_{ij}^{\bf t} = \tan\beta~m_i \delta_{ij}
-(\tan\beta+\cot\beta)V_{ti}^* V_{tj} m_j\,,
\label{e:coup}
\end{eqnarray}
where the indices $i$ and $j$ run over the down-type quarks $d,s$ and $b$, and 
the superscripts indicate which type of model we are considering, 
{\em i.e.} which quark flavor has the nontrivial transformation. 

There are tight constraints coming from neutral meson mixing, which can in principle 
go through tree-level scalar exchange. For u- and c-type models, constraints from 
such mixings force $m_H = m_A$. For t-type models, such an exact degeneracy is not needed,
in particular for low values of $\tan\beta$ \cite{Bhattacharyya:2013rya,Bhattacharyya:2014nja}, 
but as we will see, a near degeneracy
will be motivated from the LFV muon decays. 

A completely analogous formalism goes for the lepton sector. 
For massless neutrinos, one gets the coupling matrices by the following replacements: 
 $(N_u, D_u)\to 0$, $V=1$, and $N_d(D_d)\to N_e(D_e)$, so that there is 
no leptonic FCNC. For massive neutrinos, one just replaces $V$ 
by the Pontecorvo-Maki-Nakagawa-Sakata (PMNS) matrix $U$, and the $(u,d)$ labels by $(\nu,e)$. 
There can be three types of leptonic mixing models, which we will call $\nu_1$, $\nu_2$, and $\nu_3$ type
models. For example, in the $\nu_1$-type model, the coupling matrices are
\be
N_\nu = {\rm diag}\{-m_{\nu_1} \cot\beta\,, m_{\nu_2}\tan\beta\,,
m_{\nu_3}\tan\beta\}\,, \ \ 
(N_e)_{ij} = \tan\beta~m_i \delta_{ij}
-(\tan\beta+\cot\beta)U_{1i}^* U_{1j} m_j\,.
\label{Ne}
\ee
The couplings of all scalars with charged leptons are given by Eq.\ (\ref{e:coup}). 
The tree-level decay width for $H (A) \to \mu\tau$ can be written as
\begin{eqnarray}
\Gamma(H(A)\to \mu\tau) &=& \frac{1}{8\pi m_{H(A)}^3} \left[(|a|^2+|b|^2)(m_{H(A)}^2-m_\mu^2-m_\tau^2) -4m_\mu m_\tau 
\Re(ab^*) \right] \nonumber \\
&& \times \sqrt{\left\{m_{H(A)}^2-(m_\mu+m_\tau)^2 \right\}\left\{m_{H(A)}^2-(m_\mu-m_\tau)^2 \right\}}\nonumber\\
&& \approx \frac{1}{8\pi} m_{H(A)} (|a|^2+|b|^2) \,,
\end{eqnarray}
where the interaction Lagrangian is generically written as
\begin{eqnarray}
{\mathscr L}_{\rm int} = \overline{\mu} (aP_L+bP_R)\tau H(A) + {\rm h.c.}
\end{eqnarray}
The expressions for the Yukawa couplings $a$ and $b$ for $H$ and $A$ can be taken directly from Eq.\ (\ref{full-yukawa}):
\bea
{\mathscr L}_{\rm int} = -\frac{H}{v}\overline{\mu}\left[(N_e)_{\mu\tau}P_R + 
(N_e)^*_{\tau\mu}P_L\right] \tau + \frac{iA}{v}\overline{\mu}\left[(N_e)_{\mu\tau}P_R - 
(N_e)^*_{\tau\mu}P_L\right] \tau + {\rm h.c.} \,,
\eea
where the expressions for $(N_e)_{\mu\tau}$ and $(N_e)_{\tau\mu}$ can be obtained for $\nu_1$-model from \Eqn{Ne}, as
\be
(N_e)_{\mu\tau} = -\left(\tan\beta+\cot\beta\right) U^\ast_{1\mu} U_{1\tau} m_\tau\,,\ \ 
(N_e)^\ast_{\tau\mu} = -\left(\tan\beta+\cot\beta\right) U^\ast_{1\mu} U_{1\tau} m_\mu\,,
\ee
with $U_{1\mu} = -\sin\theta_{12} \cos\theta_{23}-\cos\theta_{12}\sin\theta_{23}\sin\theta_{13} e^{-i\delta}$ and $U_{1\tau} = \sin\theta_{12} \sin\theta_{23}-\cos\theta_{12}\cos\theta_{23}\sin\theta_{13} e^{-i\delta}$. The angles and phase are the same as used in the standard parametrization of the PMNS matrix.

A very tight constraint on the parameter space comes from the radiative decay $\mu\to e\gamma$. 
The amplitude for the LFV decay $\ell_i \to \ell_j\gamma$ can be written as
\begin{eqnarray}
T_\mu = \sqrt{\frac{G_F^2\alpha}{8\pi^3}}m_i \overline{\ell}_j (i\sigma_{\mu\nu}q^\nu) (C_LP_L+C_RP_R)\ell_i \,.
\end{eqnarray}
Using this, we may write the expression for the BR as~\cite{Casas:2001sr}
\begin{eqnarray}
{\rm BR}(\ell_i \to \ell_j\gamma) = \frac{3\alpha}{2\pi}\left(|C_L|^2 +|C_R|^2 \right) \,.
\end{eqnarray}
Since the charged scalar loop contribution depends on the ratio $(m_{\nu_i}^2/m_{H^+}^2)$, we neglect it on account
of tiny neutrino masses. The dominant contributions to $C_{L,R}$ will come from the neutral scalar loops mediated
by $H$ and $A$. Detailed expressions for the contributions from neutral scalar loops, for the process $b\to s\gamma$, 
already appear in the Appendix of Ref.~\cite{Bhattacharyya:2014nja}. From these, corresponding expressions for 
$\mu\to e\gamma$ can be easily obtained by straightforward replacements of the CKM elements by the appropriate
PMNS elements and the down-type masses by the corresponding masses of the charged leptons.

\begin{figure}[htbp!]
\centering
\includegraphics[scale=0.4]{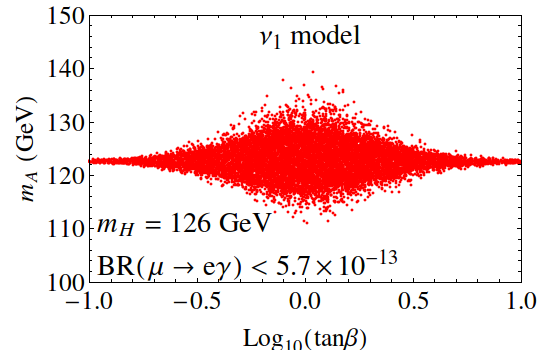} ~~
\includegraphics[scale=0.4]{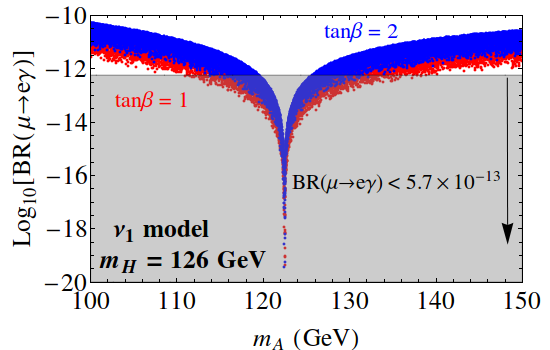}
\caption{\em (L) The allowed region in the $\tan\beta$--$m_A$ plane coming from the non-observation of 
$\mu\to e\gamma$. The horizontal width comes from an almost fine-tuned cancellation between $H$ and $A$.
(R) The Br($\mu\to e\gamma$) shows more clearly the allowed region and the cancellation. 
The shaded region is still experimentally allowed.}
\label{f:meg}
\end{figure}

\section{Analysis}

For our analysis, we will assume the BGL model to be t-type in the quark sector 
(u- and c-type models are very tightly constrained from flavor data \cite{Bhattacharyya:2014nja})
 and $\nu_1$ type 
in the lepton sector, and will call it $t\nu_1$ model for brevity (we will use the same type of 
nomenclature for other models in the BGL class). Let us first try to justify our choice. 

%
\begin{figure}[htbp!]
\centering
\includegraphics[scale=0.4]{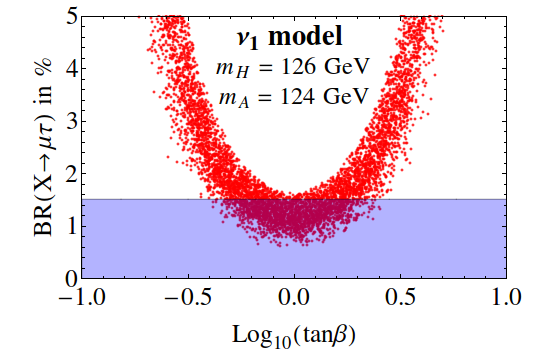}
\caption{\em The combined branching ratio for $H,A\to\mu\tau$ in the $\nu_1$ type model. The 
lower shaded region shows the 95\% upper limit for the branching ratio. In our evaluation of the BR, we have
used $\Gamma_h^{\rm SM}=4.1$~MeV as the denominator to consistently compare our results with 
the bounds given in \cite{Khachatryan:2015kon}.}
  \label{f:hmutau}
\end{figure}
%
Our starting point is the assumption that the alignment limit is exact, as far as experimental precision 
goes. In this limit, $h$ has only flavor-conserving couplings, and these 
couplings (also the gauge couplings) are 
exactly the same as those in the SM. 
Thus, the $\mu\tau$ final state, apparently originating from a 125 GeV scalar, cannot come from $h$.
We postulate that the $\mu\tau$ signal is the first hint of a quasi-degenerate nonstandard Higgs (in this
case $H$ with $m_H=126$ GeV). Under this assumption, we have checked that in the $\nu_2$ and $\nu_3$
models, the decay $H\to \mu\tau$ overshoots the experimental data for any values of $\tan\beta$, while 
for the $\nu_1$ model, the decay is found to be under control for a narrow range of $\tan\beta$. As we will discuss later, 
the experimental bound on BR($\mu\to e\gamma$) compels the pseudoscalar ($A$) to be equally light when
$m_H=126$ GeV is assumed for the $\nu_1$ model. Thus, we are forced to think of a scenario where all three 
neutral scalars of the 2HDM are quasi-degenerate at about 125 GeV. Therefore, it is important to include the contributions from
both $H$ and $A$ while comparing with the LHC signal strengths into different decay channels.
To this end, we define all the $\mu$-parameters as
\be
\mu_F = \frac{\sigma(pp \to h,H,A \to F)}{\sigma(pp \to h_{SM} \to F)}\,,
\label{def-mu}
\ee
for any final state $F$ accessible from the decay of the Higgs boson.
Of course, $H$ and $A$ can contribute to flavor-conserving fermionic final states, or even to $\gamma\gamma$. However,
in the alignment limit, they do not have couplings of the form $H(A)VV$, where $V=W,Z$ is a vector boson. 
So there will neither be any production of $H$ or $A$ through vector 
boson fusion, nor any decay to $WW^\ast$ or $ZZ^\ast$. 

Now we discuss the constraints in greater details. 
With leptonic flavor mixing, both $H$ and $A$ can contribute to LFV processes. The tightest constraint comes 
from $\mu\to e\gamma$, which is shown in Fig.~\ref{f:meg}. This shows that, barring an unnatural cancellation 
 between $H$ and $A$ contributions, the data forces $A$ to be near degenerate with $h$ and $H$. 
We have also checked that, for the allowed region from $\mu\to e\gamma$, the BRs of the processes $\tau\to e\gamma$
and $\tau\to \mu\gamma$ are always $< 10^{-12}$ which is four orders of magnitudes below the current
experimental limits \cite{Agashe:2014kda}. 
So, we have the tantalizing possibility that there may actually be three states within the 126 GeV resonance. 

As far as the decay $\mu\to 3e$ is concerned, we have checked explicitly that over the entire parameter space 
that we talk about, the branching ratio at tree-level is at most of the order of $10^{-17}$, with almost 
degenerate scalars. This is about 5 orders of magnitude smaller than the current experimental upper limit of 
$1.0\times 10^{-12}$ at 90\% CL. The loop-level decay is suppressed compared to $\mu\to e\gamma$ by at least 
one more power of $\alpha$ \cite{Harnik:2012pb}, so this process is well under control. 

Once we accept $H$ and $A$ to be degenerate with $h$, the $\mu\tau$ signal has to come from both these scalars
\footnote{For exactly degenerate states, the decay widths may change because of the interference effect \cite{Cacciapaglia:2009ic}. 
This, however, is a fine-tuned possibility which we will not enter into.}. 
In Fig.~\ref{f:hmutau} we show the branching ratio for this channel in the $\nu_1$ type model, with $m_H=126$ GeV and $m_A
= 124$~GeV. The width of the parabola-like region comes from the $3\sigma$ experimental range of the PMNS angles \cite{Gonzalez-Garcia:2014bfa}
together with the variation of the Dirac CP phase, $\delta$ in the range $[0,2\pi]$;
and a more precise determination of them will help in thinning out the parabola. The nature of the plot is 
not surprising because, as can be seen from \Eqn{e:coup}, the FCNC couplings come with a prefactor of 
$(\tan\beta+\cot\beta)$ which has a minimum at $\tb=1$. What the figure shows is that $\tan\beta$
has to be extremely constrained, $0.37 < \tan\beta < 2.8$. 
For other $\nu_i$ type models, the parabolic shape is maintained but the lowest point of the parabola lies 
well above the experimental range (this can be attributed to relatively larger magnitudes of the 
PMNS elements which are involved in the tree-level $\mu\tau$ couplings for $\nu_{2,3}$ models), which forces us to consider only the $\nu_1$ type model. It is worth noting that our scenario predicts $BR(X\to \mu\tau) > 0.5\%$ which can be tested in the next run of 
the LHC, where $X$ denotes the summed contributions from $H$ and $A$ decays.

\begin{figure}[htbp!]
\centering
\includegraphics[height=5cm,width=6cm]{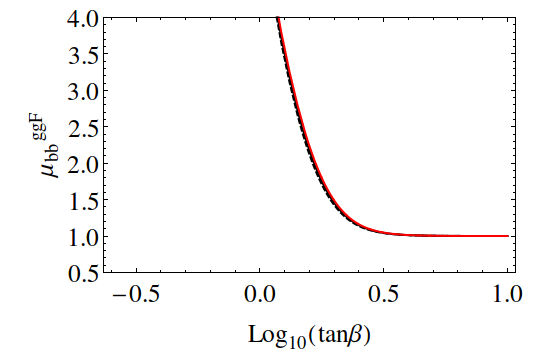} 
\includegraphics[scale=0.28]{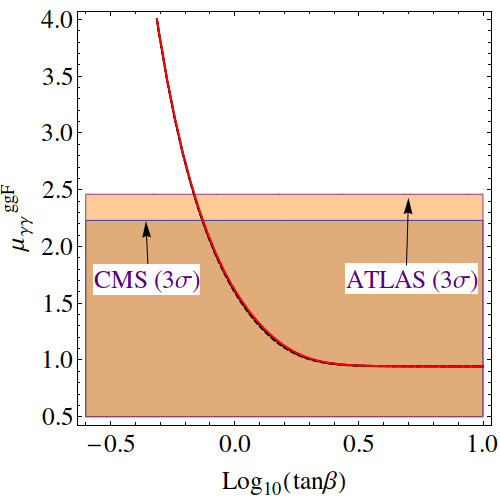}
\includegraphics[scale=0.28]{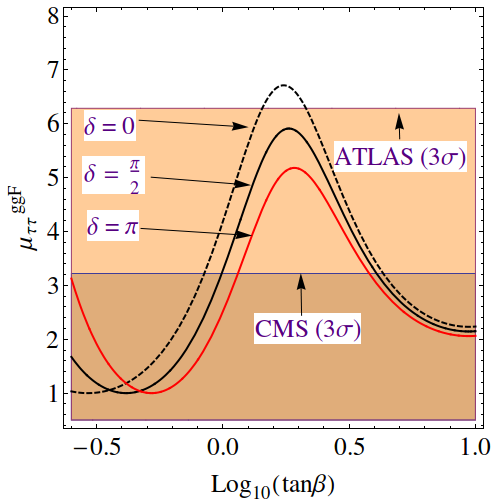}
\caption{\em $\mu$-values (for the definition, see \Eqn{def-mu}) for different final states in the ggF production 
channel, as a function of $\tan\beta$ for t$\nu_1$-model. Contributions from all the three neutral scalars ($h$, $H$ and $A$) have been included in the estimation of the signal strengths. For the plots we have assumed $m_h=125$ GeV, $m_H=126$ GeV and $m_A=124$ GeV, and the central values for various PMNS matrix elements. 
Also shown are the $3\sigma$ limits on the $\mu$-parameters from the ATLAS and CMS experiments. There are 
no significant numbers for $h\to b\bar{b}$ in the ggF channel. }
\label{f:muggF}
\end{figure}

Now that we have laid out the constraints, it is time to see why $H$ and $A$ could have been missed at the LHC. 
The only observable final states where one can have signals for these scalars are $b\bar{b}$, $\gamma\gamma$, and $\tau\tau$,
because these scalars do not have any trilinear gauge couplings.  The experimental 
numbers,  taken from Refs.\ \cite{ATLAS:tau,Khachatryan:2014jba,Aad:2014eha,Khachatryan:2014ira,Flechl:2015foa}, are, at the 
$1\sigma$ level,
\bea
\mu_{\tau\tau}^{\rm ggF} &=& 1.93\pm 1.45 \ \ {\rm (ATLAS)}\,, 0.52\pm 0.9\ \  {\rm (CMS)}\,, \nonumber\\
\mu_{\gamma\gamma}^{\rm ggF} &=&1.32\pm 0.38 \ \ {\rm (ATLAS)}\,, 1.12\pm 0.37\ \ {\rm  (CMS)}\,,
\eea
while there are no significant bounds for $\mu_{b\bar{b}}^{\rm ggF}$. 

The $\mu$-values for these three 
channels, as a function of $\tan\beta$, are shown in Fig.~\ref{f:muggF}; also shown are the $3\sigma$ bands for the 
ATLAS and CMS experiments. 
As the model is a t-type  in the quark sector, $b\bar{b}$ or even $\gamma\gamma$ constraints are relatively relaxed for $\tan\beta > 1$. 
In fact, there is no noteworthy measurement of the signal strength in the $b\bar{b}$ channel with ggF tagging 
because of the huge background. 
The $\tau\tau$ channel, on the other hand, shows a significant enhancement. Note that because of the 
nature of the PMNS coupling, the branching ratio depends on the leptonic CP-violating phase $\delta$;
the least enhancement is expected for $\delta=\pi$.

Thus, apart from the $\tau\tau$ channel, there is no reason why the new scalars should have been already seen 
at the LHC, if $\tan\beta \geq 1$. Unfortunately, we cannot go to large values of $\tan\beta$ ({\em e.g.} 10) because that 
will be in conflict with $H(A)\to\mu\tau$ data. The ATLAS data on ditau signal strength
 in the ggF channel 
is, till now, pretty inconclusive \cite{Flechl:2015foa}, and supports $\mu_{\tau\tau}^{\rm ggF}$ as large as 5. 
The CMS data \cite{Khachatryan:2014jba} is much more precise (although the ggF and the VBF
channels are not always differentiated very successfully\cite{Chatrchyan:2014nva}). Anyway, all the relevant $\mu$-values are 
well within the experimental $3\sigma$ range; if we take the range at the $2\sigma$ level, $\mu_{\tau\tau}^{\rm ggF}
\approx 2.3$, which forces the parameter space to a region near $\tan\beta=1$ and large nonzero $\delta$.
We find $\mu_{\gamma\gamma}^{\rm ggF}\approx 1.5$ for $\tan\beta \approx 1$, which is still within the 
$2\sigma$ allowed range of both ATLAS and CMS \cite{Flechl:2015foa,Khachatryan:2014jba}. 
This is another prediction which can be tested in precision Higgs studies. 

\begin{figure}[htbp!]
\begin{minipage}{0.46\textwidth}
\centerline{\includegraphics[width=8cm,height=6cm]{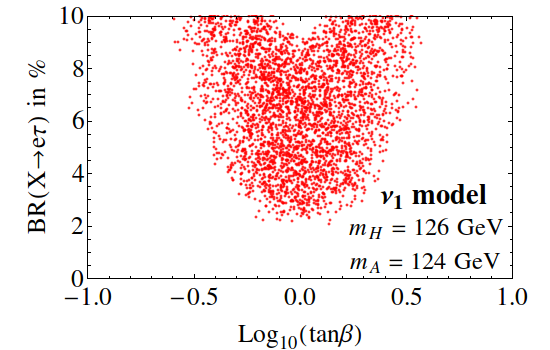}}
\caption{\em Prediction for the branching ratio in the $e\tau$ channel. The PMNS 
angles have been varied within their $3\sigma$ experimental range.}
\label{etau}
\end{minipage}
\hfill
\begin{minipage}{0.46\textwidth}
\centerline{\includegraphics[width=8cm,height=6cm]{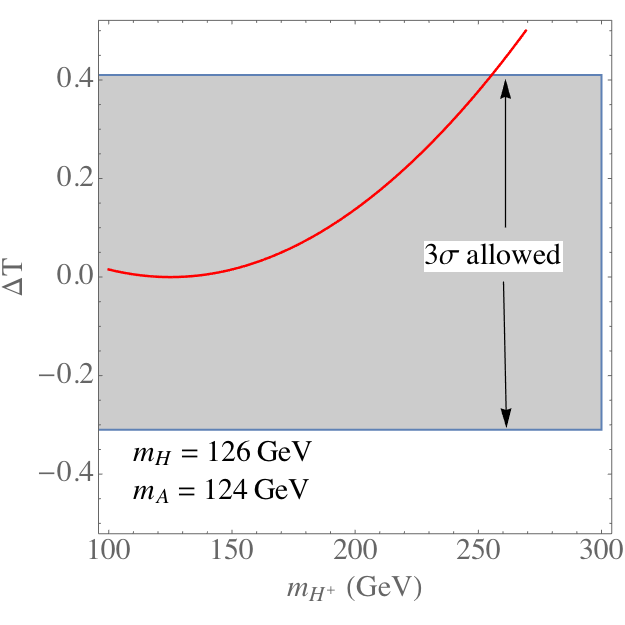}}
\caption{\em Constraint on $m_{H^+}$ at 99\% C.L. from the experimental limit on $\Delta T$.}
\label{t-parameter}
\end{minipage}
\end{figure}

In passing, we note that while finding a similar excess in the $e\mu$ channel will be tough due to the smallness
of the masses of the leptons involved, our scenario does predict a branching ratio for the $e\tau$ channel 
to be above 2\% (see Fig.~\ref{etau}), which can
be used, among other things, to falsify the model. 

The charged scalar merits a brief discussion. 
With nearly degenerate $H$ and $A$,  the strongest indirect bound on the charged 
scalar mass ($m_{H^+}$), for the BGL model at hand, comes from the oblique $T$-parameter.
In the 2HDM alignment limit, the new physics contribution 
to the $T$-parameter can be expressed as \cite{He:2001tp,Grimus:2007if}
\begin{eqnarray}
\Delta T &=&  \frac{1}{16\pi\sin^2\theta_wM_W^2} 
\left[{\mathcal{F}}(m_{H^+}^2,m_H^2) + 
\mathcal{F}(m_{H^+}^2,m_A^2) - \mathcal{F}(m_H^2,m_A^2) \right] \,, \\
{\rm with,}~~~~
\mathcal{F}(x,y)&=&\left\{
\begin{array}{cc}
\frac{x+y}{2} - \frac{xy}{x-y}\ln\left(\frac{x}{y} \right) & {\rm for}~~ x\neq y \,, \\
0 & ~{\rm for}~~x=y \,.
\end{array}
\right.
 \label{tparam}
\end{eqnarray}
Taking the new physics contribution to the $T$-parameter as \cite{Baak:2013ppa}
\begin{eqnarray} 
\Delta T = 0.05 \pm 0.12 \,,
\end{eqnarray}
we show the $3\sigma$ range for the allowed charged scalar mass in this model 
in Fig.~\ref{t-parameter}. We find that the value of $m_{H^+}$  can rise as high as 250 GeV.
Since the LHC is most sensitive for $m_{H^+}$ up to 160 GeV \cite{CMS:2014cdp,Aad:2014kga},
a charged scalar heavier than that can remain well hidden from the current reaches of LHC.

\section{Summary}

We have shown that the recently observed excess in the $\mu\tau$ channel coming out of the 126 GeV resonance 
can be interpreted in a radically different way than the canonical explanation of having one more scalar with 
LFV couplings which mixes with the SM doublet and the one of the resultant CP-even neutral mass eigenstates
showing the LFV decay signal due to the slight admixture. We took a 2HDM with nonzero FCNC, and note that 
this model can also be tuned to the canonical explanation mentioned above to generate the LFV decays, with 
hardly any originality. Taking a different approach, we assumed that the alignment limit is exact, and explore 
the possible consequences. This is how our logic went.

\begin{itemize}
\item In the alignment limit, the SM Higgs boson $h$ cannot have any tree-level FCNC coupling. The $\mu\tau$ 
signal, therefore, must be coming from another scalar, which is almost degenerate with $h$. The splitting has to 
be less than the resolution for two nearby mass peaks. 

\item The 2HDM of Branco, Grimus, and Lavoura has at least two neutral scalars with tree-level FCNC, one being the 
CP-even $H$ and the other, CP-odd $A$. We, therefore, assume $H$ to be nearly degenerate with $h$. 
\item This poses a serious problem with the non-observation of $\mu\to e\gamma$. This process receives 
contribution from both $H$ and $A$ mediated diagrams, but the amplitudes come with opposite signs. The way out 
is to take $A$ to be nearly degenerate with $H$, so that the contributions more or less cancel. 

\item $H$ and $A$ cannot be differentiated at the LHC. Thus, we have to take both $H\to\mu\tau$ and $A\to\mu\tau$ 
into consideration while calculating the branching ratio. Due to the nature of the coupling, the combined branching ratio 
has a minimum at $\tan\beta=1$ and grows on either side. For $\nu_2$ and $\nu_3$ type models, the lowest 
possible value of the branching ratio is way above the data, so we are forced to consider the $\nu_1$ model.
This gives an allowed range of $\tan\beta$ as $0.4 < \tan\beta < 2.8$. If we take a t-type model, all constraints 
coming from hadronic physics are satisfied in this range of $\tan\beta$. 

\item $H$ and $A$ do not have any trilinear gauge couplings in the alignment limit. Therefore, the only way to produce them 
at the LHC is through gluon fusion (they can be radiated off a top, though). In the gluon fusion channel, there is 
no significant constraint for $b\bar{b}$ final states. The $\gamma\gamma$ final states are significantly enhanced by 
top loops for $\tan\beta < 1$ but is well under control for $\tan\beta \geq 1$. The only important channel is 
$\tau\tau$, which receives a significant enhancement. However, with the present state of the data, signatures for 
the new scalars in these channels can still be below any statistical significance.   

\item Most importantly, the model is testable and falsifiable through various means. First, a more precise determination 
of the branching ratio in the $\mu\tau$ channel can rule out this model, if the upper limit drops below the 
lowest possible value of the branching ratio. Note that the model has no free parameters once one specifies 
$\tan\beta$ (the neutral scalars are all nearly degenerate and the charged scalar is not relevant here; however,
with $m_H \approx m_A$ in the alignment limit, $m_{H^+}$ cannot be too far away from $m_h$ for the theory to be consistent 
with the electroweak precision observables, in particular the $T$-parameter 
\footnote{The couplings are such that there is no constraint from $b\to s\gamma$ even for such low-mass $H^+$ \cite{Bhattacharyya:2014nja}, but direct searches by ATLAS and CMS prefer the mass range of 160-180 GeV.}
\cite{Bhattacharyya:2013rya}) and therefore 
no way to play with the minimum branching ratio. Second, the model predicts a large branching ratio for $\mu\to 
e\gamma$, almost close to the experimental upper limit, which should be seen in the next generation of 
experiments, unless there is an unnatural cancellation between two competing amplitudes. Third, the model 
predicts a large branching ratio for the $e\tau$ final state, whose lowest value is about 2\%. Fourth, and the most 
clinching, is a precise determination of $h\to\tau\tau$ in the gluon fusion channel. If there is hardly any scope for 
a large enhancement, this model is ruled out. 
\end{itemize}

\centerline{\bf Acknowledgements}
D.D.\  thanks Maria Hoffman and Atanu Modak for many helpful discussions. D.D. also thanks the Department of Atomic Energy, India for financial support. A.K.\ would like to acknowledge Department of Science 
and Technology, Government of India, and Council for Scientific and Industrial Research, Government of India, for 
support through research projects.


\bibliographystyle{JHEP}
\bibliography{Ref.bib}  
\end{document}